
 \input harvmac

\def \om {\omega}

\def\const {{\rm const}}
\def \s {\sigma}
\def\t {\tau}

\def \p {\phi}
\def \ha {\half}
\def \ov {\over}

\def \four{{\textstyle {1\ov 4}}}
\def \a {\alpha}
\def \lr { \lref}
\def\ep{\epsilon}

\def\vp {\varphi}
\def \bd {\bar \del}
\def \r {\rho}
\def\const {{\rm const}}\def\bd {\bar \del} \def\m{\mu}\def\n {\nu}\def\l
{\lambda}

\def\g {\gamma}

\def\y {{ \tilde y}}

 \def \sm {$\s$-model\ }
\def   \td {\tilde }

\def \lr { \lref}

\gdef \jnl#1, #2, #3, 1#4#5#6{ { #1~}{ #2} (1#4#5#6) #3}

\lr \orb {L. Dixon, J. Harvey, C. Vafa and E. Witten, \np B274 (1986) 285.}

\lr \burg{C.P. Burgess, \np B294 (1987) 427; V.V. Nesterenko, \ijmp A4 (1989)
2627.}
\lr \tsenul {A.A. Tseytlin, \np B390 (1993) 153. }
\lr \susskind { L. Susskind, ``Some speculations about black hole entropy in
string theory", RU-93-44 (1993), hep-th/9309135. }

\lr \dab {A. Dabholkar, \jnl \np, B439, 650, 1995;
D.A. Lowe and A.  Strominger, \jnl \pr, D51, 1793, 1995.}

\lr\hrt {G.T. Horowitz and A.A. Tseytlin. \pr D50 (1994) 5204. }

\lr \gtwo {E. Del Giudice, P. Di Vecchia and S. Fubini, Ann. Phys. 70
(1972) 378; K. A. Friedman and C. Rosenzweig, Nuovo Cimento 10A (1972) 53;
S. Matsuda and T. Saido, Phys. Lett. B43 (1973) 123; M. Ademollo {\it et al},
Nuovo Cimento A21 (1974) 77;
S. Ferrara, M. Porrati and V.L. Teledgi, Phys. Rev. D46
(1992) 3529.}

\lr \plane { D. Amati and C. Klim\v c\'\i k,
\jnl \pl, B219, 443, 1989;
 G. Horowitz and A. Steif,  \jnl \prl, 64, 260, 1990; \jnl \pr,
D42, 1950, 1990;
 G. Horowitz, in: {\it
 Strings '90}, eds. R Arnowitt et. al.
 (World Scientific, Singapore, 1991);
 H. de Vega and N. S\' anchez, \pr D45 (1992) 2783; Class. Quant. Grav. 10
(1993) 2007.}

\lr \gibma { G.W.  Gibbons and  K. Maeda, \np B298 (1988) 741.}
\lr\gib{G.W.  Gibbons, in: {\it Fields and Geometry}, Proceedings of the 22nd
Karpacz
Winter School of Theoretical Physics, ed. A. Jadczyk (World Scientific,
Singapore,  1986).}

\lref \tsnul { A. Tseytlin, \jnl \np, B390, 153, 1993.}

\lr\duval  { C. Duval, Z. Horvath and P.A. Horvathy, \jnl \pl,  B313, 10,
1993.}

\lr\gro{D.J.  Gross, J.A. Harvey,  E. Martinec and R. Rohm, \np B256 (1985)
253; \np B267 (1985) 75.}
\lr \duff {Duff et al Nepomechie }
\lr \incer { E.J. Ferrer, E.S. Fradkin and V.de la Incera, \pl B248 (1990)
281.}
\lr \green {M.B. Green, J.H.  Schwarz and E.  Witten, {\it Superstring Theory}
(Cambridge U.P., 1988).}

\lr \quev {C. Burgess and F. Quevedo,  \np B 421 (1994) 373. }
  \lr \nahm { W. Nahm, \np B124 (1977) 121. }

 \lr \kok {  E. Kiritsis and C. Kounnas,  ``Infrared behavior of closed
superstrings in strong magnetic and gravitational fields", hep-th/9508078. }

\lr \sen{A. Sen, \pr D32 (1985) 2102; \prl 55 (1985) 1846.}
\lr \hulw { C. M. Hull and E.  Witten,  \jnl \pl, B160, 398, 1985. }

\lr\hult { C. Hull and P. Townsend, \jnl \pl, B178, 187, 1986. }
\lr \gps {S.  Giddings, J. Polchinski and A. Strominger, \jnl  \pr,  D48,
 5784, 1993. }

\lr \ghr { S. Gates, C. Hull and M. Ro\v cek, \np B248 (1984) 15.}
\lr \jon {C. Johnson, \pr D50 (1994) 4032.}
\lr \landau {L.D.  Landau and E.M.  Lifshitz, {\it Quantum Mechanics} (Pergamon
Press, N.Y., 1977).  }
\lref \horts {G.T. Horowitz and A.A. Tseytlin, \pr  D51 (1995) 2896. }

\lr \attick {J.J. Atick  and E. Witten, \np B310 (1988) 291. }
\lr\fund{A. Dabholkar, G. Gibbons, J. Harvey and F. Ruiz Ruiz, \jnl \np, B340,
33, 1990;
D. Garfinkle, \jnl \pr, D46, 4286, 1992; A. Sen, \jnl \np, B388, 457, 1992;
D. Waldram, \jnl \pr, D47, 2528, 1993. }

\lr\planetach{ J. Garriga and E. Verdaguer, Phys. Rev. {\bf D43} (1991) 391.}

\lr\rut{J.G. Russo and A.A. Tseytlin, {``Exactly solvable string models
of curved space-time backgrounds"},
 \np B449 (1995) 91, hep-th/9502038.}
\lr\melvint{A.A. Tseytlin,  \pl B346 (1995) 55. }
\lr\rts{J.G. Russo and A.A. Tseytlin, \np B448 (1995) 293,
hep-th/9411099.}
 \lr \gross {D.J.  Gross, J.A. Harvey,  E. Martinec and R. Rohm, \np B256
(1985)
253; \np B267 (1985) 75.}

\lr \gso{
M.B. Green, J.H.  Schwarz and E. Witten, {\it Superstring Theory} (Cambridge
U.P., 1987). }

\lr \sen{A. Sen, \pr D32 (1985) 2102; \prl 55 (1985) 1846.}
\lr \hulw { C. M. Hull and E.  Witten,  \jnl \pl, B160, 398, 1985. }

\def\np {  Nucl. Phys. }
\def \pl { Phys. Lett. }

\def \prl { Phys. Rev. Lett. }
\def \pr  { Phys. Rev. }

\def \ijmp { Int. J. Mod. Phys. }

\baselineskip8pt
\Title{\vbox
{\baselineskip 6pt{\hbox{CERN-TH/95-215}}{\hbox
{Imperial/TP/94-95/57}}{\hbox{hep-th/9508068}} {\hbox{
   }}} }
{\vbox{\centerline {Magnetic flux tube models  in superstring theory}
 }}

\vskip -30 true pt

\centerline  { {J.G. Russo\footnote {$^*$} {e-mail address:
 jrusso@vxcern.cern.ch
} }}

 \smallskip \smallskip

\centerline{\it  Theory Division, CERN}
\smallskip

\centerline{\it  CH-1211  Geneva 23, Switzerland}

\medskip
\centerline {and}
\medskip
\centerline{   A.A. Tseytlin\footnote{$^{\star}$}{\baselineskip8pt
e-mail address: tseytlin@ic.ac.uk}\footnote{$^{\dagger}$}{\baselineskip8pt
On leave  from Lebedev  Physics
Institute, Moscow.} }

\smallskip\smallskip
\centerline {\it  Theoretical Physics Group, Blackett Laboratory,}
\smallskip

\centerline {\it  Imperial College,  London SW7 2BZ, U.K. }
\bigskip
\centerline {\bf Abstract}
\medskip
\baselineskip10pt
\noindent
Superstring models describing curved 4-dimensional
 magnetic flux tube backgrounds
are exactly solvable in terms of free fields.
We first consider the simplest model of this type
(corresponding to `Kaluza-Klein' $a=\sqrt 3$ Melvin   background).
Its  2d action has a  flat but topologically non-trivial 10-dimensional
target  space (there is a  mixing  of  angular coordinate of the 2-plane  with
an internal compact coordinate).
We demonstrate that this theory has broken supersymmetry but  is perturbatively
 stable if the  radius  $R$ of  the internal  coordinate is larger than
$R_0=\sqrt{2\a'}$.
In the  Green-Schwarz formulation
 the supersymmetry breaking is a consequence of the presence
of a flat but non-trivial  connection
in the fermionic terms in  the action.
For $R< R_0$ and the magnetic field  strength parameter $q >R/2\a'$, there
appear instabilities  corresponding to   tachyonic winding  states. The torus
partition function  $Z(q,R)$ is
finite for $R>R_0$ and vanishes for $qR=2n$\  ($n=$integer).
At the special points $qR=2n$  ($2n+1$)  the model
is equivalent to the free  superstring theory compactified on a circle
with periodic (antiperiodic)
 boundary condition for space-time fermions.
Analogous results are  obtained for a more general class of static magnetic
flux tube geometries including the $a=1$ Melvin model.

\medskip

\Date {August 1995}

\noblackbox
\baselineskip 14pt plus 2pt minus 2pt
\def \X { {\cal X}}
\vfill\eject
\def \N {{\hat N}}
 \lr\mel {M.A.  Melvin, \pl 8 (1964) 65. }

\lr \tser { A.A. Tseytlin, ``Exact solutions of closed string theory",
hep-th/9505052. }
\lr \gaun {F. Dowker, J.P. Gauntlett, G.W. Gibbons and G.T. Horowitz, ``The
decay of magnetic fields in Kaluza-Klein theory", hep-th/9507143.}
\lr\dowo{  F. Dowker, J.P. Gauntlett, D.A. Kastor and J. Traschen,
\pr D49 (1994) 2909; F. Dowker, J.P. Gauntlett, S.B. Giddings and G.T.
Horowitz, \pr D50 (1994)
2662.}

\lr \ruh{J.G. Russo and A.A. Tseytlin, ``Heterotic strings in a uniform
magnetic field", \np  B454 (1995) 164, hep-th/9506071.}

\lr\carl {S. Carlip, \pl B186 (1987) 141;
R.E.  Kallosh and A. Morozov, \ijmp A3 (1988) 1943.}
%
\lr \wit {E. Witten, \np B195 (1982) 481.}
\def \L {\Lambda}

\lr \hul {G.W. Gibbons  and C.M. Hull, \pl B109 (1982) 190;
C.M. Hull, \pl B139 (1984) 39;
R. G\" uven, \pl B191 (1987) 241.}
\lr \kallosh { E.A.  Bergshoeff,  R. Kallosh and T. Ort\'\i n, \pr D47 (1993)
5444.  }
\lr \grs { M.B. Green and J.H. Schwarz,  B136 (1984) 307; \np B243 (1984) 285.
}

\lr \witte{ E. Witten, \np B266 (1986) 245;
M.T. Grisaru, P. Howe, L. Mezincescu, B.~Nilsson and P.K.  Townsend,
\pl B162 (1985) 116;
J.J. Atick, A. Dhar and B.~Ratra, \pl B169 (1986) 54. }
\lr \fra{E.S. Fradkin and A.A. Tseytlin, \pl B160 (1985) 69.}

\lr\banks{T. Banks, M. Dine, H. Dijkstra and W. Fischler, \pl B212 (1988) 45;
I.~Antoniadis, C. Bachas and A. Sagnotti, \pl B235 (1990) 255;
 J. Harvey and J. Liu, \pl B268 (1991) 40;
R. Khuri,  \pl B294 (1992) 325; \np B387 (1992)  315; J. Gauntlett, J. Harvey
and J. Liu,
 \np  B409 (1993)  363;  S.  Giddings, J. Polchinski and A. Strominger,  \pr
D48
 (1993)  5784;
C.~Johnson,  \pr D50 (1994)  4032:
C. Bachas and E. Kiritsis, \pl B325 (1994) 103. }

\lr\bach{ C. Bachas, ``A way to break supersymmetry", hep-th/9503030.}

\lr \kiri {E. Kiritsis and  C. Kounnas, \np  B442 (1995) 472;
 ``Infrared regulated string theory and loop corrections to coupling
constants", hep-th/9507051. }

\lr \alvi{L. Alvarez-Gaum\' e, G. Moore and C. Vafa, {Commun. Math. Phys.} 106
(1986) 1. }

\lr \rohm {R. Rohm, \np B237 (1984) 553.}
\lr \bri {M.B. Green, J.H. Schwarz and L. Brink, \np B198 (1982) 474.}

\lr\teee{A.A. Tseytlin, ``Closed superstrings in magnetic field: instabilities and supersymmetry breaking", 
 to appear in: {\it ``String Gravity and Physics at the Planck Scale",} 
 Proceedings of  Chalonge School, Erice, 8-19 September 1995,  ed. N. S\'anchez (Kluwer, Dordrecht), hep-th/9510041.}   
\lr \barny{I. Bars, D. Nemeschansky and S. Yankielowicz, \np B278 (1986) 632. }
\lr \cand{P. Candelas, G. Horowitz, A. Strominger and E. Witten, \np B258 (1985)  46.}

\lr \deal{S. de Alwis, J. Polchinski and R. Schimmrigk, \pl B218 (1989) 449.}

\lr\sche{ J. Scherk and J.H. Schwarz,
 Phys. Lett. { B82} (1979) 60; Nucl. Phys. {B153} (1979) 61.}
\lr\sch{S. Ferrara, C. Kounnas and M. Porrati, \np B304 (1988) 500;
 Phys. Lett.  { B197} (1987) 135; Phys. Lett. { B206} (1988) 25;
C. Kounnas and M. Porrati, Nucl. Phys. { B310} (1988) 355;
I. Antoniadis, C. Bachas, D. Lewellen and T. Tomaras, \pl B207 (1988) 441.}
\lr \koun{V.P. Nair, A. Shapere, A. Strominger and F. Wilczek, \np B287 (1987)
402; P. Ginsparg and C. Vafa, \np B289 (1987) 414; H. Itoyama and T.R. Taylor,
Phys. Lett. { B186} (1987) 129.}

\lr \att{J.J. Atick and E. Witten, \np B310 (1988) 291.}
\lr \kou {C. Kounnas and B. Rostant, \np B341 (1990) 641.}
\lr\diens{K.R. Dienes, \np B429 (1994) 533; hep-th/9409114;
hep-th/9505194.}
\lr \bank {T. Banks and L. Dixon, \np B307 (1988) 93.}

 \def \h {\chi}
\def \da {\del_a}
\def \daa {\del^a}
\def \d {\del }

\def \y {{\td y}}
\def \b {\beta}
\def \S {{\cal S}}

\newsec{Introduction}
Magnetic backgrounds
were actively studied  recently  from various points of view
 in the context of both field theory and string theory
(see,  e.g.,  \refs{\mel,\gib,\gibma,\dowo,\gaun}
and \refs{\banks,\kiri,\bach,\rts,\melvint,\rut,\ruh}).
Of particular interest are  the simplest  ones -- static flux tube
type configurations  with approximately uniform magnetic field  generalizing
the Melvin solution.
Such backgrounds are exact solutions of string theory \refs{\melvint,\rut}
and, moreover, the spectrum of the corresponding conformal string models  can
be explicitly determined \rut. In the bosonic case these
theories are generically unstable due to the appearance of tachyons for certain
values of the magnetic field parameters.\foot{In addition to
this perturbative instability there may be other instabilities
 of non-perturbative
origin, discussed in the field-theory framework,  in \gaun.}

 The problem addressed in the  present  paper is the construction and  solution
of the corresponding superstring versions.
 We shall find that  there still exists a range of parameters
for which  magnetic flux tube backgrounds  considered as
solutions of superstring theory are perturbatively
unstable.

In Section 2 we shall review the structure of the bosonic string model
which represents  a particular  ($a=\sqrt 3$) Melvin flux tube background in
$D=4$.

The corresponding type II superstring theory will be solved  in
Section 3 using RNS formulation. Its  quantum Hamiltonian  will be
the free superstring one plus terms linear and quadratic in
 angular momentum operators.
As a result, the mass spectrum can be explicitly determined.

The basic properties of the  spectrum
will be studied in Section 4.
We  will show that supersymmetry is   broken and that there  exist
intervals of values of moduli parameters
(Kaluza-Klein radius and magnetic field strength)
for which the model is unstable.
We shall also discuss  the heterotic version of the model.

In Section 5 we shall consider the light-cone Green-Schwarz
 formulation of the
theory, which turns out to be very simple.
The breaking  of supersymmetry will be related to the
absence of Killing spinors in the   Melvin background.
We shall also compute the expression for the
partition function on the torus which will be finite  or infinite depending on
the values of the parameters.

In Section 6 the results of  Sections 3-5,
 obtained for the $a=\sqrt 3$ Melvin model,
will be generalized to  a class of static magnetic flux tube  models
which includes, in particular, the $a=1$ Melvin model.
We shall explain the reason for solvability of these models
and clarify the nature of perturbative
instabilities that appear for generic values  of the magnetic
field  parameters.

Section 7  will contain a summary and  remarks on some  generalizations.
In particular,   we shall comment on the
 relation  between  the $a=\sqrt {3}$ Melvin  model
and the  superstring compactifications on twisted tori
where supersymmetry is broken by discrete twist angles  (or by the
`Scherk-Schwarz' mechanism).

\newsec{Bosonic string model for $a=\sqrt 3$ Melvin background }

In this and the following two sections  we  shall
consider the supersymmetric version of the simplest representative in the class
of   static
magnetic flux tube models  of ref. \rut\ --  the `Kaluza-Klein' (or $a=\sqrt
3$) Melvin model. It  has
 properties similar to those of the more general models but yet can be solved
in a rather simple  way.
 This theory  is special in that the corresponding \sm has  flat  target space
 of non-trivial topology (other models in \rut\  have  curved  target spaces
but are related to flat models by angular duality and globally non-trivial
coordinate shifts).
 The relation of $a=\sqrt 3$  Melvin  background \refs{\mel,\gibma}
  to flat $D=5$ theory was  pointed out
at the field-theory level in  {\dowo}  (see also  \gaun)
and at the string-theory level in  \refs{\rut,\tser}.\foot{The  string model
corresponding to the $a=1$  Melvin background \gibma\  was  constructed  in
\melvint\ and solved in \rut. In contrast to the flat 3-space geometry
of the $a=\sqrt 3$ model (see below), in the $a=1$ case
the 3-space  is curved  and the  $(\r,\vp)$-surface
asymptotically closes at large $\r$.}
The bosonic string model is defined by the following Lagrangian
\eqn\mode{ L=L_0 + L_1 \ , \ \ \ \ \ \
L_0 = -\d_a t \d^a t + \d_a x_\a \d^a x^\a \ , \ \ \ \ }
\eqn\mee{
L_1 = \d_a \r \d^a \r + \r^2 (\d_a \vp + q \da y)(\daa \vp + q \daa y )  + \da
y \daa y  \ . }
Here $\r\geq 0 $ and $0 < \vp \leq 2\pi $  correspond to cylindrical
coordinates  on a ($x_1,x_2)$-plane, $y$ is a  circular `Kaluza-Klein'
coordinate
with period $2\pi R$, and $x^\a $ include the  flat $x^3$-coordinate
of $D=4$ space-time and, e.g., 21  (or 5 in  the superstring case)  internal
coordinates compactified on a torus.

The constant $q$ plays the role of the magnetic field strength parameter in the
4-dimensional  interpretation. $L_1$ can be represented in the  `Kaluza-Klein'
form
 \eqn\me{
L_1 = \d_a \r \d^a \r +  F(\r) \r^2 \d_a \vp \daa \vp
  +  e^{2\s}  ( \da y   +  {\cal A}_\vp \da \vp)   ( \daa y   +  {\cal A_\vp}
\daa \vp)    \ ,  }
so that the $D=4$ background (metric, Abelian vector  field $\cal A$ and
scalar $\s$) corresponding to \mode\ is indeed the $a=\sqrt 3$ Melvin geometry
\eqn\back{ ds^2_4=  - dt^2  + d\r^2 +   F(\r)  \r^2
d\vp^2 + dx_3^2 \ , }
\eqn\baa{ {\cal A}_\vp=  q  F(\r)   \r^2 \ , \ \ \  \ e^{2\s} = { F}\inv
(\r) \ , \ \ \  F  \equiv {1\ov 1 +  q^2 \r^2}  \  .    }
The non-trivial 3-dimensional part \mee\ of \mode\  is non-chiral (there is no
antisymmetric tensor background) and the dilaton is constant. The 3-metric
\eqn\met{ds^2 = d\r^2 + \r^2 (d\vp +q dy)^2 + dy^2 \ ,
 }
is  flat  (so that the model is  automatically conformal to all orders)
since  locally one may
introduce the coordinate $\theta= \vp + q y$ and decouple $y$ from
$\r, \vp$.
The global structure of this 3-space is non-trivial:
the fixed $\r$ section is a 2-torus (with $\r$-dependent conformal factor and
complex modulus) which degenerates into a circle at $\r=0$.  The
space  is  actually regular everywhere, including  $\r=0$ (this can easily be
seen by rewriting \met\  in
terms of Cartesian coordinates
of the 2-plane  and $y$, cf. (2.7) below).
 It can also be     obtained
by factorizing $R^3$ over the group generated by
 translations in  two angular directions:
in the coordinates where $ds^2=
d\r^2 + \r^2 d\theta^2 + dy^2 \ (\theta= \vp + q y)$ one should
identify the points $(\r,\theta, y) = (\r, \theta + 2\pi n   + 2\pi qRm, y+
2\pi R m)$ \ ($n,m=$integers), i.e. combine the shift  by $2\pi R$ in $y$ with
a rotation by an arbitrary angle $2\pi qR$ in the 2-plane.\foot{Since the
orbits of this group are non-compact (in contrast to, e.g.,  the case of a
special 2-cone$= R^2/Z_N$ orbifold \refs{\orb,\dab})
the  corresponding string  model  can be defined for arbitrary  continuous
values of the moduli parameters $q, R$.}
Although the space is flat, the  corresponding string theory will  be
non-trivial
(already at the classical level due to the existence of winding string states
and at the quantum level
in the non-winding sector where there will be  a `magnetic' coupling
to the  total angular momentum in the 2-plane),
representing an   example of a gravitational
 Aharonov-Bohm-type  phenomenon: the value of the parameter  $q$ does not
influence the (zero) curvature of the space but affects the  global properties
like masses of string states.\foot{Let us  also note that the \sm which is
$\vp$-dual  to $L_1$ is
a special case of   models  in \rut,
$\  \td L_1 =   \d_a \r \d^a \r + \r^{-2} \d_a \td \vp \daa \td \vp
+ q \ep^{ab} \da y \d_b \td \vp
 + \da y \daa y  +  \a' R^{(2)} (\p_0 - \ln \r)  . $
The constant torsion term corresponds in $D=4$   to the `Aharonov-Bohm' gauge
potential
$B_{iy}\equiv {\cal B}_i = q \ep_{ij} x^j/x^2$,
$\ F({\cal B})_{ij} = - 2\pi q \ep_{ij} \delta^{(2)} (x) $.}

 The new coordinate $\theta$
is globally defined ($2\pi$ periodic)  only
for special integer  periods of $qy$, i.e. for
$ qR = n, \  \ n= 0, \pm 1 , ... \ .$
In these cases  \me\ is trivial, i.e.  equivalent to  a free  bosonic string
theory compactified on a circle.\foot{The Kaluza-Klein  field theory
is  also trivial in this case, since
the corresponding solution of $D=5$ Einstein  theory is
equivalent to (Minkowski  4-space)$\times S^1$
(any  `observable'  computed  in terms of the 4-dimensional  variables, i.e.
on the background \back,\baa,
should  have the same value  as in the $D=5$ theory).}
Models with $n <qR< n+1$  are equivalent to  models with  $0 < qR <1$.
We shall see that this periodicity condition in $qR$  will be  modified in
superstring
theory: because of the presence of fermions
of half-integer spin,  $n$  will be replaced by  $2n$, i.e.   only  models
with $qR=2n$ will be trivial.
More generally,  superstring theories with $(R,q)$  and $(R, q + 2n R\inv)$
will be equivalent.

It should be noted  that it is $e^\s R$ in (2.3),(2.5)  that  plays the role of an effective Kaluza-Klein radius of the compact  fifth dimension.
Since $e^\s$  grows with radial distance from the flux tube, 
it may seem that the Kaluza-Klein interpretation eventually breaks down
(as was discussed in the field-theory context in \gaun\ the  standard 
Kaluza-Klein interpretation would apply only for $qR <<1$).
As for the  higher dimensional string theory, it  
is  defined for an arbitrary $q$. Its mass spectrum  derived below
will not contain  extra light states (in the non-winding sector) for small $R$
(and arbitrary $q$). Still, 
 its 4-dimensional `magnetic' interpretation 
will  directly apply only for small $qR$.

The model \mode, \mee\   has   a straightforward generalization  where  $\vp$
is `mixed' with several  internal coordinates:
$\d_a \vp + q \da y \to \d_a \vp + q_r \da y^r$, etc.
The corresponding $D=4$ background  contains several magnetic  fields and
moduli fields.

It is useful to represent \me\ in the following equivalent form,
introducing $x= x_1 + i x_2 = \r e^{i\vp} $ :
 \eqn\mes{
L_1 = (\d_a x_i - q \ep _{ij} x_j \d_a y) (\daa x_i - q \ep _{ij} x_j \daa y) +
\da y \daa y  }
$$
 = (\da x  + i q x\da y   ) (\daa x^*  - i q x^*\daa y   )+ \da y \daa y $$
\eqn\cov{  = D^a x D^*_a x^* + \da y \daa y   \ ,  \ \ \ \  D_a \equiv \da + iq
\da y \ ; }
  we therefore get a  charged  complex  2d scalar field $x$
in a flat 2d gauge potential. Since $y$ is compact,
the effect of this gauge potential  will be non-trivial.

The conformal theory   corresponding to the bosonic model \mode\
was solved in  \rut\ (as a special case of a more general class of models)
by  observing    that the solution  of the classical string equations can be
expressed in terms of free fields and applying
 the canonical quantization.
This procedure   is particularly simple in the present
model  \mes.
 Since  the  `$U(1)$  potential'  $q \da y$ in \cov\  is flat,   $x$  can be
formally `rotated' to  decouple $x$ from $y$.
Then $y$ satisfies the free field equation and $x$ is  also expressed in terms
of free fields.
 The only   interaction   which  effectively survives in the final expressions
is the coupling of $x$ to the derivative of the zero mode part of $y$,
$\ y_*= y_0 + 2\a' p \t + 2Rw\s\ $.
It is then straightforward to  carry out the canonical quantization procedure,
expressing all observables in terms of free oscillators.
The resulting  Hamiltonian is given by the sum
of the free string  Hamiltonian plus  $O(q)$ and $O(q^2)$ terms
depending on the left and right components of the free string angular momentum
operators $\hat J_L$ and $\hat J_R$ \rut.

As was shown  in \rut, this  bosonic string  model
 is stable in the
non-winding  sector, where there  are  no new  instabilities in
addition to the usual flat space tachyon. This means, in particular,  that
the  Kaluza-Klein field  theory corresponding to   the Melvin  background is
perturbatively stable with respect to the `massless' (graviton, vector,
scalar)  {\it and} massive perturbations. This theory
may still be unstable  at a non-perturbative  level \gaun.
At the same time,
 there exists a range of parameters $q$ and $ R $
for which  there  are tachyonic states in the winding  sector, i.e. this
string model is unstable
against certain winding-mode perturbations.

This instability  (whose origin is  essentially in
 the gyromagnetic coupling  term
$wq R(\hat J_R -\hat J_L)$, which may have negative  sign) is not related to
the presence of the flat
bosonic string tachyon
and may thus  be expected to survive  (for certain values of $q$ and $R$)
also
in the superstring case.
This, indeed, is  what will be found below.

\newsec{Solution of  the superstring  Melvin model}
In what follows we shall  consider
the  type II superstring version of \mode\  (heterotic models
with the magnetic field in the Kaluza-Klein sector can be obtained by
straightforward  `left' or `right' truncations and    have similar
properties).
Since
 (in contrast to the  constant magnetic field model in \refs{\rts,\ruh}) the
$\del y$-dependent interaction terms in  \mee\  are  non-chiral,  there does
not exist
an associated heterotic string model with the magnetic field embedded in the
internal gauge sector.

In this section we shall  consider the   RNS formulation of the model.  The
model can be solved also  by using  directly
the Green-Schwarz  \refs{\grs,\gso}  formulation (see Section 5), which
 confirms (and clarifies certain aspects of)  the  RNS  solution.
The $(1,1)$  world-sheet supersymmetric extension
of the  model \mee,\cov\ has the form ($x^\m\equiv (x^i, y)$)
\eqn\suu { L_{\rm RNS} = G_{\m\n} (x)\del_+  x^\m \del_-  x^\n    } $$+ \
\l_{R m} (\delta ^m_n \del_+    + \om^m_{\ n \m }\del_+ x^\m ) \l^n_R +
 \l_{L m} (\delta ^m_n \del_-    + \om^m_{\ n \m }\del_- x^\m ) \l^n_L \ . $$
$\l^m= e^m_\m \l^\m$ are  vierbein components of the 2d
Majorana-Weyl spinors and $\om^m_{\ n \m }$ is the (flat)   spin   connection.
There are  no quartic fermionic terms since the metric is flat.
In the natural  basis $e^i= dx^i - q\ep^{ij} x_j dy, \ \ e^y= dy$, the spin
connection 1-form  has the following components
\eqn\spi{
\ \om^{ij} =- q\ep^{ij}dy \ , \ \ \ \ \ \ \om^{iy} =0
 . }
In terms of the left and right   Weyl spinors $\l=\l_1 +i\l_2 $ corresponding
to $x=x_1 +ix_2$ and $\l^y\equiv  \psi$, we get (cf. \cov)
\eqn\onn{L_{\rm RNS}=   D_+ x  D^*_- x^* +
\d_+ y \d_- y
  +  \l^*_R  D_+\l_R
 +  \l^*_L D_-  \l_L  }
$$
+ \  \psi_R \d_+ \psi_R +   \psi_L \d_- \psi_L \ , \ \ \ \ \
 D_\pm \equiv  \del_\pm   +  iq \del_\pm  y\ ,     $$
 where the covariant derivative $ D_\pm  $
is the same as in \cov, i.e.  it contains the  flat $U(1)$ potential.
 This means that, as in the bosonic case,
 it is possible to redefine the fields $x$, $\l$
 so that the only non-trivial   coupling  that will remain at the end
will be to the zero mode of $y$.\foot{
One can directly generalize the bosonic case discussion by replacing $x^i,y$
by $(1,1)$ superfields and  observing  that the zero mode  part
of the $y$-superfield can have  only the bosonic component $y_*$.
Note that both the  momentum  and  the winding
parts of $y_*$  are on an equal footing in \onn:
the model is  non-trivial (not equivalent to the  free string one)
 already  in the non-winding sector.}
Although it may seem that, as in the bosonic case,  the model  with
 $qR=n$
should be  equivalent to the free superstring theory compactified on a circle
(since for $qR=n$
 one can,    in principle,   eliminate the coupling terms in \onn\ by rotating
the fields)
this will not actually be true    unless the integer $n$
 is even,  $n=2k$.
The non-triviality  for $n=2k+1$
is directly related to the presence of space-time
fermions in the spectrum,  which change sign  under $2\pi $ spatial  rotation
accompanying the periodic shift in $y$ (see below and Section 5).

Taking the world-sheet to be  a  cylinder $(\tau, \s)\ $
($ 0 < \s \leq \pi) $  we can solve the classical equations corresponding to
\onn\ by introducing  the     fields  $X$  and $\L_{R,L}$,
which  will satisfy the free string equations  but  will have
`twisted' boundary conditions ($\s_\pm \equiv \t\pm \s $)\foot{The twist
parameter $\g$  can be interpreted as a flux
corresponding to the 2d $U(1)$ field $A_a = q \del_a y$ on the cylinder,
$\int A= 2qRw\int d\s=2\pi \g$.
Note that we have redefined $\gamma$ by factor of 2 compared to our
previous papers \refs{\rut,\ruh}.}
\eqn\bix { x(\t,\s) = e^{-iq y(\t,\s)} X (\t, \s) \ ,   \ \ \ \
\d_+\d_- X = 0 \ , \ \ \ X= X_+ (\s_+) + X_- (\s_-) \ , }
\eqn\bii{  \ \  X(\t,\s + \pi ) = e^{2\pi i \g  }X(\t,\s) \ , \ \
\ \ \ \g \equiv  qRw\ ,  }
\eqn\red{ \l _{R,L}(\t,\s) = e^{-i q y(\t,\s)  } \L_{R,L} (\t,\s)\ ,
\ \ \ \   \  \d_\pm \L _{R,L} =0 \ ,  \ \ \
\L _{R,L} = \L _{R,L} (\s_\mp)  \ , }
\eqn\bou{
 \L _{R,L }(\t,\s + \pi ) = \pm e^{2\pi i  \g   } \L_{R,L} (\t,\s)\ ,   }
 with the signs `$\pm$' in \bou\
corresponding to the Ramond (R)  and Neveu-Schwarz (NS)  sectors.
The crucial observation is that $y$  still satisfies
the free-field equation:
\eqn\yyyi{ \d_+\d_- y =0 \ , \ \ \ y= y_* + y'\ , \ \ \ y_*= y_0 + p_+ \s_+
+ p_- \s_-\ . }
We have used  the fact that the  fields $x,y,\l$  must obey the usual
closed-string boundary conditions,
\eqn \gam{ x(\t,\s + \pi ) = x(\t,\s)\ , \ \ \ \  \
 y(\t,\s + \pi ) = y(\t,\s) + 2\pi R w \ , \ \ \ \ w = 0, \pm 1 , ... \ ,  }
\eqn\fer{ \l_{R,L} (\t, \s+  \pi) = \pm \l_{R,L} (\t,\s)\ . }
The  explicit expressions for the fields $X=X_+ + X_-$ and $\L_{L,R}$  are then
\eqn\bcfz{
 X_\pm (\s_\pm)  = e^{\pm 2i \g \s_\pm } \X_\pm  (\s_\pm) \ ,  \ \ \ \ \
\X_\pm
(\s_\pm \pm \pi)=
\X_\pm (\s_\pm )\ , }
\eqn\laa{ \L _{L,R}(\s_\pm ) = e^{\pm  2i \g \s_\pm   } \eta_{L,R} (\s_\pm)\ ,
}
where $\X_\pm $  and $\eta_{L,R}$
are the free  fields with the standard free closed string
boundary conditions, i.e.
\eqn\fourie{ \X_+ =  i  \sqrt{\ha \a' } \sum_{n \in {\bf Z}} \tilde a_n e^{-2in
\s_+}   \ ,
 \ \   \   \X_- =  i  \sqrt{\ha \a' }
\sum_{n \in {\bf Z}} a_n  e^{-2in \s_-}  \ , }
\eqn\fourram{
 \eta_{R}^{(\rm NS)}=\sqrt{2\a'}\sum_{r\in {\bf Z}+\ha } c_r\
e^{-2ir\s_- }\ , \ \ \ \ \eta_{R}^{(\rm R)} =\sqrt{2\a'}\sum_{n \in {\bf Z}}
d_n\
e^{-2in\s_-}\ ,
}
and similar expressions for  the left fermions with oscillators having extra
tildes.
We can then proceed with canonical quantization
of the model expressing the observables in terms of the above free oscillators.
It is convenient  to choose   the  light-cone gauge,
eliminating  the oscillator part of $u=y-t$ (see  \refs{\rut,\ruh} for
details).
Then
\eqn\uuu{ u=u_*\equiv u_0 + 2\a' (p + E)\t  +  2Rw \s   \ , }
$$
p= p_y -q\hat J\ , \ \ \ \ \ p_y= mR\inv \ , \ \ \ \ m=0, \pm 1 , ... \ ,  $$
where $E$ is the total energy, $m$ is the Kaluza-Klein linear momentum number,
$w$ is the winding number and
$\hat J=\hat J_R + \hat J_L $ is the total angular momentum in the 2-plane.

In what follows we shall first assume that $w$ (or $qR$)
is such that $0\leq \g <1$ and then consider generalizations to other values of
$\g=qRw$.  The angular momentum operators that appear in the  final
Hamiltonian contain the orbital momentum parts plus the  spin parts (with the
latter having  the  standard free superstring form  \gso)
 \eqn\angulr{
{\hat  J}_R= - b^{\dagger }_0 b_0 -\ha    +\sum_{n=1}^\infty \big(  b^{\dagger
}_{n+}b_{n+} - b^{\dagger }_{n-}  b_{n-} \big)+\hat K_R  \ , }
$$
 \hat  K^{\rm (NS)}_R=-  \sum_{r=\ha }^\infty (c_{r}^* c_{r} + c_{-r} c_{-r}^*)
 , \ \
 \ \hat  K^{\rm (R)}_R=-[d_0^*,d_0]  +
  \sum_{n=1}^\infty (d_{n}^* d_{n}+d_{-n}  d_{-n}^*)
 .   $$
The expression of $\hat J_L$ is similar,   with  the reversed sign of the
orbital momentum terms.
Here  $b$'s are  the  free  creation and annihilation operators related to the
modes in \fourie\ by rescaling  by factors $(n \pm  \g)^{1/2}$, see \rut.
The eigenvalues of $\hat J_{L,R}$  are
\eqn\eig{  \hat J_{L,R}
 = \pm (l_{L,R} + \ha) + S_{L,R} \   ,  \ \
 \hat J\equiv \hat J_L + \hat J_R = l_L-l_R + S_L + S_R,    }
where  the orbital momenta $l_{L,R}=0,1,2,...  $ (which replace the continuous
linear momenta $p_1,p_2$ in the 2-plane
for non-zero values of $\g$)
 are  the analogues of the Landau quantum number
 and $S_{R,L}$ are the  spin components.\foot{In the case $\gamma=0$ (or, more
generally,  $\g=n$)
  the zero-mode structure changes in that the translational invariance
in the 2-plane is recovered, see \refs {\rts , \rut }. This  leads to  a slight
modification in the formulas (the operators $b_0^\dagger, b_0, \tilde
b_0^\dagger, \tilde b_0$   in the expressions below  are then  replaced by
standard zero-mode operators $x_{1,2}, p_{1,2}$). We shall not explicitly
indicate this in what follows.}

The  number of states operators $\hat N_R$ and $\hat N_L$
have the standard form
\eqn\coo{ \hat N_{R,L}= N_{R,L} -a \ ,
\ \  \
\ \ \    a^{\rm (R)} =0\ , \ \ \    a^{\rm (NS) } =\ha \ , }
where,   e.g.   in the Ramond sector,
\eqn\nnn{
 N^{\rm (R)}_R= \sum_{n=1}^\infty n \big( b^{\dagger }_{n+}b_{n+}+ b^{\dagger
}_{n-}b_{n-}
+ b^{\dagger }_{n\a} b_{n\a}  +  d^*_nd_n+d _{- n}d_{-n}^* + d _{-n\a}
d_{n\a}\big) \ .  }
 $ N^{\rm (R)}_L$ has a similar expression
in terms of  operators with tildes (there are no  contributions with
oscillators corresponding to $y$ and $t$ since we used the light-cone gauge).
Under the  usual GSO projection
(which is necessary for the correspondence  with the Green-Schwarz formulation
and
 with the free RNS  superstring theory in the limit $q=0$ but will not imply
the  space-time  supersymmetry in the present case)
  $\N_{R}$  and $\N_{L }$ can    take  only non-negative integer values
(and correspond to  the number of states operators of  the light-cone
Green-Schwarz formulation).

The resulting expressions for the Hamiltonian and level matching constraint
are\foot{Symmetrizing the classical expressions
for the Virasoro operators $L_0, \tilde L_0$,
we   then normal-order them and  use  the  generalized $\zeta$-function
prescription.  In contrast to the bosonic case \rut\
here the $\g^2$ normal ordering terms cancel out
between bosons and fermions.
For example, in
the NS-sector one obtains: \   $L_0\to $  $ L_0 -\ha (1-  \g )  $ .
The latter normal-ordering constants are  naturally absorbed into $\hat
N_{R,L}$ and $\hat J_{R,L}$.}
\eqn\hamil{
\hat H =
 \ha  \a' \big( -E^2 +  p_\a^2 + \ha Q_L^2 +
\ha Q_R^2  \big) + \hat N_R+  \hat N_L
\  }
$$ -
\a' q (  Q_L  \hat J_R  + Q_R  { \hat J}_L)    + \ha \a' q^2 \hat J^2   \ ,  $$
\eqn\defi{  Q_{L,R} \equiv  {m\ov R} \pm  {wR\ov \a'}\ ,  }
\eqn\cons{\N_R-  \N_L = mw  \ .  }
$\hat H$ can be represented also in the following  (`free superstring
compactified on a circle') form
which  clarifies its structure and is useful for generalizations
\eqn\ham{
\hat H =
 \ha  \a' \big( -E^2 +  p_\a^2  + {m'}^{2}R^{-2}  +  {\a'}^{-2}  w^2R^2 \big)
+ \hat N_R'+  \hat N_L'
\ ,   }
where $m', \hat N_R',  \hat N_L'$ (which are no longer integer in general) are
defined by
\eqn\bbb{  m'\equiv m - qR \hat J \ ,  }
\eqn \kkk{  \hat N_R' \equiv  \hat N_R  - \g  \hat J_R\ ,
\ \ \ \   \hat N_L' \equiv  \hat N_L  +  \g  \hat J_L\ , \ \ \ \ \ \ \g=
qRw\ . }
Up to the orbital momentum terms,  $ \hat N_{R,L}'$ can be put into  the same
form  as free operators $ \hat N_{R,L}$  (see \nnn)
 with  the factor $n$ replaced by $n \pm  \g$.

The Virasoro  condition $\hat H=0$  leads to
the following  mass spectrum
\eqn\kme{
M^2 \equiv E^2 -p_\a^2 =M^2_0 - 2qR\inv m \hat J
-2{\a'}\inv q  R w (\hat J_R -\hat J_L) + q^2 \hat J^2\ ,
}
where $M_0$ is the mass operator of the free superstring compactified on a
circle (for simplicity we ignore the contributions of  the other 5
free compactified dimensions, i.e. the  corresponding  momenta are set equal to
 zero)
\eqn\kmee{
 M^2_0 =2{\a'}\inv (\N_L+ \N_R)   +   m^2R^{-2}
 +   {\a'}^{-2}  w^2R^2  \ .
 }
The equivalent form of \kme , which demonstrates that in general $M^2$ is not
positive-definite in the winding sector,  is
 \eqn\ano{ M^2 =
2{\a'}\inv (\N_R+ \N_L)   +
(mR\inv  -  q  \hat J)^2 } $$
 +  \   [{\a'}\inv wR -
q (\hat J_R -\hat J_L)]^2  -   q^2    (\hat J_R -\hat J_L)^2 \ .   $$
Let us consider first  the  zero winding sector $w=0$ ($\g=0$)
 where
$ M^2 =
2{\a'}\inv (\N_R+ \N_L)   + { m'}^{2} R^{-2}$.
It is clear from \bbb\ that the mass spectrum is then invariant under
\eqn\inve{  q\to q+ 2nR\inv\ , \  \ \ \ \ n=0, \pm 1 , ...\ , }
  since this transformation can be compensated by
$m\to m- 2n\hat J=$ integer.  Note that  since $\hat J$ can take  both  integer
(NS-NS, R-R sectors) {  and}
{\it half-integer}  (NS-R, R-NS sectors) values, the symmetry of the bosonic
part of the spectrum  $q\to q+ nR\inv$  is {\it not}
 a   symmetry of  its
fermionic part, i.e. the full  superstring spectrum is invariant only under
\inve.

The same conclusion about the periodicity in $q$  is  true in  general for
$w\not=0$.    In the form given  above,
eqs. \hamil,\kme,\ano\  are valid  for   $0\leq w<(qR)\inv $, i.e. for
$0\leq \g < 1$. The generalization to other values of
  $\g $, e.g.  $\g $ in any   interval $n\leq \g <n+1 $, $n=$integer,
 is straightforward (see also \rut). The net effect is the  replacement of $
\g$ in \kkk\  by $ \g-n $,  i.e.
$qRw$ in  \kme\ by $qRw-n$. The general form of the mass operator is thus
\eqn\masgr{
\a' M^2 = 2 (\N_R+ \N_L)   +
\a' (mR\inv  -  q  \hat J)^2  + {\a'}^{-1}  w^2R^2}
$$    -\     (\g -n) (\hat J_R -\hat J_L)
\  .
$$
As   will be clear from a comparison with the Green-Schwarz
formulation, one should use the standard  GSO projection  for $2k\leq
\g<2k+1$, and  the  `reversed' one   for
$2k +1 \leq  \g<2k + 2 $, \ ($k=0,\pm 1, ...$).  The `reversal' of GSO for $2k
+1 \leq  \g<2k + 2 $ implies  that in this interval
 only states having half-integer eigenvalues
of the operators $\hat N_{L,R}$ will survive, including, in particular,
scalar odd-winding  tachyon states with  $\hat N_L=\hat N_R=-\ha $.
 This prescription (which
appeared  also in the model  of  \rohm, see also
 \refs{\att, \kou}, 
related to the special case $qR=2n +1$ of our model) 
is consistent with the modular invariance of the partition
function (see Section 5).

For fixed radius $R$ the     mass spectrum  is thus periodic in $q$,
  i.e. it is  mapped into itself under \inve\
(combined with $m\to m- 2n\hat J$).
In the case $qR =2n$ (i.e.  $\g=2nw=2k$) the
 spectrum is thus equivalent to the  standard spectrum of the free superstring
theory compactified on a circle.
For $qR= 2n+1$ (i.e.  $\g=(2n +1)w=2k +1$ if $w$ is odd)
the spectrum is the same as that
of  free superstring compactified
on a circle with antiperiodic boundary conditions
for space-time fermions \rohm \  (see
also \refs{\att,\kou}).
 This relation will become clear
 in
the Green-Schwarz formulation (Section 5).
 In particular, it will be apparent that
the interaction term in the  superstring action
 can be eliminated by a globally defined field transformation
only  if  $qR=2n$, while  for $qR=2n +1$
this can be done at the expense of imposing
antiperiodic boundary conditions (in the $y$-direction)
 on fermions (under the rotation by
the angle $2\pi qR= 2\pi$  in  the 2-plane,
 which is associated with a periodic shift in $y$,  the bosons remain invariant
but the spinors  change sign).

Let us note that,  in contrast to the case  of the
free string compactified on a circle,
the mass spectrum \ano\  (for generic $qR$)
is {\it not}  invariant under the naive duality transformation
$R \to \a'R\inv $ (accompanied by some redefinition of quantum numbers such as
$ (w,m) \to (m,w)$ in the free string case).
Unlike, e.g. the free string or  $a=1$ Melvin model \refs{\melvint,\rut},
the  action  \mee\ does not preserve its form  under the duality transformation
in $y$,  i.e.   the
$y$-duality maps \mee\  into a  {\it different}
 $\s$-model (belonging to  the  3-parameter  class of models in
 \rut)
\eqn\duaa{\td L= \del_+ \r \del_-\r +  F(\r)\r^2 (\del_+ \vp  + q \del_+ \td y)
( \del_-\vp
-q \del_-  \td y) + \del_+ \td y \del_- \td y  } $$
+ \   {\cal R} (\p_0 +  \ha \ln F ) , \ \  \ \ \ \  F\equiv (1 + q^2 \r^2)\inv
\ ,  \ \ \ \ \ {\cal R}\equiv \four \a' \sqrt{g} R^{(2)} \ .  $$
This model
is  equivalent to \mee\ at the CFT level,  i.e.  it has, in particular,
  the same  mass spectrum   \ano.

\newsec{Mass spectrum: supersymmetry breaking and (in)stability}
There are two  immediate   consequences   that can be drawn
from the above  expressions \kme,\ano\  for $M^2$:

\noindent
(i) the space-time supersymmetry is broken for $qR\not=2n$;

\noindent
(ii) there  exists a range of values of parameters  $q$ and $R$  for which
there are tachyonic states in the spectrum.

Suppose  that we start with the free superstring compactified on a circle $y$
 and study
what happens with the spectrum when we switch on the
magnetic field,    $q\not=0$.
Since the mass shift in \kme\ involves  {\it both }
components  $ \hat J_L$ and $ \hat J_R$  of the angular  momentum
(with independent generically non-vanishing coefficients),
it is easy to see  that  the masses of bosons and fermions that were equal for
$q=0$  will become different for $q\not=0$.
Indeed, it is impossible to have  both $\hat J_L$ and $\hat J_R$   equal for
bosons and fermions.\foot{In the constant magnetic model \ruh\
 where the coupling to the magnetic field was only through  $\hat J_R$
 (half of) the supersymmetry was preserved in the  type II  superstring
and in the   `left-right' symmetric and `left' heterotic   models. There
$\hat J_R$ (and thus the  mass shift)
was the
same for  bosons and fermions.}

Supersymmetry is absent  already   in the { non-winding}
 sector (where the coupling  is to  the total angular momentum  $\hat J$). For
example,
the free superstring  massless  (ground) states
($\hat N_{L,R}=0=m=w$)  will, according to \kme,    get masses
 $M=|q\hat J|$  proportional to their total angular  momenta,
which   must be integer for bosons and half-integer
 for fermions (cf. \eig).
Note that these
 states are neutral,  so that  from  the
 4-dimensional point of view the shift in the masses
can be interpreted as a gravitational effect.
This  shift implies,
in particular,  that supersymmetry is  broken  at   the field-theory (e.g.,
$D=5$ supergravity) level, in agreement
with the absence of Killing spinors in  the  Melvin background  (see  Section
5).

In the absence of supersymmetry  some  instabilities of the bosonic string
model may survive also  in the superstring case. As in the bosonic case, the
mass operator \ano\ is positive in the non-winding  sector, but   tachyonic
states   may   appear  in the winding sector (we use the name `tachyon'
for a state with $M^2 <0$; it should be remembered, of course,  that
 the string states we are
discussing propagate in curved $D=4$ space-time). 
 Consider, for
 example, the  NS-NS   superstring  winding states  with zero  Kaluza-Klein
momentum and  zero orbital momentum  quantum numbers    and with
maximal absolute values of spins $S_{R,L}$ at given levels
\eqn\sta{ w>0\ , \ \ \  m=0\ , \ \ \ l_R=l_L=0 \ , \ \  \
 S_R=\N_R +1 \ , \ \ \ S_L = -\N_L -1 \ . }
We will restrict our consideration to  states for which
 $0< qRw <1$ (states with  $w > (qR)\inv $  can be analysed in a similar way,
see \masgr).
Then \cons,\kme\ imply
\eqn\imp{ \N_R=\N_L\equiv N \ , \ \ \  \ \  \hat J= 0\ , \ \  \
\ \   \hat J_R-\hat J_L= 2N + 1 \ , }
\eqn\anot{ \a'M^2 =
4 N  +     {\a'}^{-1}  w^2R^2
 - 2 qR w (2N + 1  ) \ .   }
A state  with given $N$ and $w$ will be tachyonic  for  $q > q_{\rm cr},$
\eqn\crit{
q_{\rm cr}=  { 4 N    +   {\a'}^{-1}w^2R^2 \ov 2(2N +1)wR }\  .
}
For $N=0$ we get $\a' q_{\rm cr} = \ha wR$.
  The condition $qRw<1$ is satisfied provided
   $wR < \sqrt {2\a'}$.

 In general, it is easy to check  (using the fact that
$-\hat N_{R,L}-1\leq S_{R,L}\leq \hat N_{R,L}+1 $)  that
states with $M^2<0$ can be present only  for $R<\sqrt{2\a' }$, i.e.   the full
spectrum  is tachyon-free if $R>\sqrt{2\a'}$.
For fixed $R<\sqrt{2\a' }$  the minimal value of the magnetic field  strength
parameter at which  tachyons  first appear is
$\a' q_{\rm cr} = \ha R$, corresponding  to the $N=0, w=1$ case  of \crit.
Tachyons with  $N=1$ are found   at  larger values of the magnetic field $q>
q_{\rm cr}$ with  $ q_{\rm cr}$  given by \crit\ for $N=1$, etc.

All other  sectors (R-R, R-NS, NS-R)  are tachyon-free.
 Let us consider, for example,  the  fermionic  states  of the
   R-NS -sector  with the following quantum numbers (cf. \sta):
$w>0,\  l_R=l_L=0,\  S_L=-\hat N_L-1, \ S_R=\hat N_R + \ha $.
For $qRw <1$  the corresponding mass formula is (cf. \anot)
\eqn\mass{ \a'M^2 =
2(\hat N_R + \hat N_L) (1-qRw)   +    {\a'}^{-1}  (wR - \ha  {\a'} q)^2    }
$$ +\ \a' R^{-2}m(1-qRw)[m(1-qRw) + qR] \ , $$
 i.e.  $M^2$ is non-negative. The winding  fermionic
 state with  $\hat N_{R,L} =0=m, \  w=1$ thus
becomes  massless  at $q= 2R/\a'$.

Some
 values of the radius, such as      $R=\sqrt {2\a '}$, are special.
For $R=\sqrt {2\a '}$  the  value of $M^2$ is  non-negative. As the magnetic
field  $q$ is gradually
increased from zero, the masses of  the infinite number of modes
  belonging to  the set   \sta\ with $w=1$
will  decrease. They will  simultaneously
   approach zero   when $q$
will approach   $R\inv = 1/\sqrt {2\a '}$\  ($qRw\to 1$).
At this  point there is a discontinuity in $M^2$  since,  for
 $qR=1$,  the
spectrum  is equivalent
to that of a free  superstring  on a circle
 with antiperiodic boundary conditions for fermions (see
the discussion on the periodicity of $M^2$ in   $q $
in the previous section).

 The  structure of the spectrum  of the present model is  thus
different from  that of  the constant magnetic field   model
\refs{\rts,\ruh}  in which
infinitely many instabilities appeared for any arbitrarily small value of
the magnetic field.

One can consider also the heterotic version of the above model
(where the magnetic field is embedded in the Kaluza-Klein sector)
by combining the `left' or `right' part of the superstring model
with the  free internal part.  The mass formula and
the level matching  condition
 in this case take
the following  form  (cf. \ano,\cons)
\eqn\anoh{ \a' M^2 =
2  (\N_R+ \N_L+ \ha p^2_I)   +
\a' (mR\inv  -  q  \hat J)^2 } $$
 +  \   {\a'}\inv (wR)^2
- 2qRw (\hat J_R -\hat J_L)   ,    $$
$$ \N_R  -\N_L = mw + \ha p^2_I  \ , \ \ \ \  \N_R= 0,1,2, ... , \ \ \   \N_L=
N_L -1 = -1, 0, 1, ... \ , $$
where $\hat N_L$  contains only the free internal  oscillator modes (see  \ruh\
 for notation).
There are instabilities similar to the  ones discussed in the above type II
model. In addition,
there are  other instabilities   which, in the case of the special  `self-dual'
value of the  radius $R=\sqrt{\a' }$, appear  for  infinitesimal values of the
magnetic field. These are just the usual Yang-Mills-type  magnetic
instabilities, associated with the   gauge bosons  (with quantum numbers
$m=w=\pm 1$, $p_I^2=l_R=l_L=0$, $\hat N_R=N_L=0$, $S_R= 1$, $S_L=0 $)
of the $SU(2)_L$ group.

\newsec{Green-Schwarz formulation  and partition function }
The supersymmetry  breaking  is related to   the
coupling of fermions in \onn\ to
the flat but  globally non-trivial  $U(1)$ connection.
This  can be  seen   explicitly  in the Green-Schwarz formulation
\refs{\grs,\gso}
where the absence of supersymmetry is  connected  to the non-existence
of Killing spinors in a given bosonic background.
Let us  consider  the  Killing spinor equation
\eqn\kil{ (\del_\m + \four \om^{mn} _{\ \ \m} \g_{mn} ) \ep =0 \ , }
 in
the $D=3$ background corresponding to \mes.
Here $\ep= \ep (x^i,y)$ is  a space-time  spinor and
$\om^{mn} _{\ \ \m}$ is the same flat spin connection as in \suu,\spi\ so that
\kil\ reduces to
\eqn\kiii{  (\del_y - \four q \ep^{ij}\g_{ij} ) \ep =0 \   . }
The formal solution of \kiii\
\eqn\kii { \ep (y) = \exp (\four q \ep^{ij}\g_{ij} y)\ \ep (0) \ ,  }
does not, however, satisfy the
periodic boundary condition in $y$,  $\ \ep (y + 2\pi R)= \ep (y)$
(unless $qR=2n$   when  the Killing spinor does exist,  in agreement with the
fact that in this case the theory is   equivalent to the free
superstring).\foot{Redefining $\vp\to \vp - q t $
(which is always possible since $t$ is non-compact)
one can   put the  Lagrangian  \mode,\mes\  in the `plane-wave' form (see
\rut)
$\ L= \da u \daa v  + qx^ix_i \da u \daa u +
2 q\ep _{ij} x^i \da x^j \daa u  + \da x_i \daa x^i   , \ \ u= y-t, \ v= y + t
. $
Then  the absence of supersymmetry in the  Melvin model seems
to contradict usual   claims  that plane-wave  backgrounds
are  supersymmetric  (see, e.g., \refs{\hul, \kallosh}).
In fact, there is no contradiction since the supersymmetry  may be  broken in
the plane-wave backgrounds if
the direction  $y$ in which the wave is propagating is compact.
 If the  spin connection
has constant $y$  (or $u$) component,
the corresponding Killing spinor equation  may
 not have solutions consistent with periodic boundary conditions in the
$y$-direction.}

The  conclusion is that  for $qR\not=2n$ there is no residual space-time
supersymmetry
in the higher-dimensional (e.g., $D=5$ supergravity)
 counterpart of the $a=\sqrt 3$ Melvin background. The absence of Killing
spinors in the case of the $a=0$ Melvin solution of the Einstein-Maxwell theory
was previously mentioned in \gib.


Given a  generic  curved  bosonic background,  the  corresponding
Green-Schwarz (GS) superstring action  \grs\   defines  a complicated
non-linear  2d theory (see,  e.g.,   \refs{\cand,\fra,\witte}).
This theory
 appears to be  more tractable  when one is able to fix a light-cone
gauge (in particular, when the background is flat  at least in  one time-like
and one space-like  directions). Then the  action takes a `$\s$-model' form,
which  can be explicitly  determined  \fra,  e.g.   by comparing with  the
known light-cone superstring vertex operators \gso.
   This   light-cone  gauge action  becomes  very simple (quadratic in
fermions)   when the  background  geometry is flat as in the   case of
the Melvin model \mee\ (cf. \suu)
\eqn\gss{ L_{GS}
= G_{\m\n} (x)\del_+  x^\m \del_-  x^\n
+  i \S_R {\cal D}_+ \S_R
 +    i \S_L {\cal D}_- \S_L\ , } $$  {\cal D}_a\equiv \del_a  + \four
\om^{mn}_{ \m }\g_{mn}\del_a x^\m
\ .  $$
 Here $S^p_{R,L}$ $\ (p=1,...,8)$ are   the right and left  real spinors of
$SO(8)$ (we consider type IIA theory).
In the  case of \mee\ we get  (cf. \cov,\onn,\kiii)\foot{ Light-cone gauge may
be fixed on $t-x_3$.
An equivalent approach   is first  to redefine the fields to eliminate the
`oscillating'
part of $y$ from the interaction terms and then impose the  gauge on $u=t-y$.}
\eqn\ongs{L_{GS} =   (\del_+ + i q\d_+ y )x (\del_- - i q\d_- y ) x^* +
\d_+ y \d_- y  } $$
  + \   i \S_R (\del_+ - \four q \ep^{ij}\g_{ij}\d_+ y ) \S_R
 +  i \S_L (\del_- - \four q \ep^{ij}\g_{ij}\d_- y ) \S_L
\ .  $$
It is
 natural  to decompose  the $SO(8)$ spinors according
to $SO(8) \to SU(4) \times U(1)$,  i.e.  $\ \S^p_L \to (\S^r_L, \bar \S^r_L)$,
$\ \S^p_R \to (\S^r_R, \bar \S^r_R)$, \ $r=1,..,4$ ($\bar \S_{L,R}$ are complex
conjugates of $\S_{L,R}$).
 With respect to  the rotational group $U(1)$ of the plane,
$\S^r_R,  \bar \S^r_L$  and $\bar \S^r_R,   \S^r_L$ have  the  charges $\ha$
and $-\ha$  (the bosonic  fields $x,x^*$ have  the charges $\pm 1 $).
 Then the fermionic terms in \ongs\ become
\eqn\jjj{ L_{GS} (\S_{L,R})= i \bar \S^{r}_R (\del_+ + \ha iq \d_+ y ) \S^r_R +
i \bar \S^{r}_L (\del_- - \ha iq \d_- y ) \S^r_L \ . }
 The condition that the action \gss,\ongs\ has  residual supersymmetry
invariance  $\S \to \S + \ep(x)$ is  equivalent to
${\cal D}_a  \ep (x(\t,\s))=
\del_a x^\m (\del_\m  + \four  \om^{mn}_{ \m }\g_{mn}) \ep (x) =0$.
The absence of   supersymmetry  invariance is the  consequence of the absence
of  zero modes of  the  above covariant derivative operators,   or,
equivalently,  of the non-existence of  solutions of the Killing
spinor equations \kil, \kiii.

The connection terms in the covariant derivatives in the fermionic part of the
GS action \jjj\ have extra coefficients $\ha$ with respect  to the ones in the
RNS action \onn. This immediately implies that the full   theory is periodic
under $qR \to qR + 2n$.

As in the bosonic and RNS cases,   we can   explicitly  solve
the classical
string equations corresponding to \ongs\ with the final result that
  the only  essential
difference,  as compared to the free superstring case,
is the coupling of bosons and fermions to the  zero-mode part
of the flat $U(1)$ connection $\del_a y_*$.
The  expressions for the superstring Hamiltonian and mass spectrum are
essentially the same as
in \hamil , \masgr , where, for $2k\leq  \gamma<2k+1$
the  operators $\hat N_{L,R}$, $\hat J_{L,R}$  have  the usual  free GS
superstring  form,  which is  similar to their form  in the R-sector of the RNS
formalism with vanishing zero-point energy.
For  $2k-1\leq  \gamma<2k$ the operators  $\hat N_{L,R}$ have
 the `NS-sector' form,  i.e.
they  take half-integer eigenvalues starting from
$-\ha$.

The reason for  this  change from integer to half-integer eigenvalues
can be understood directly from the action \jjj:
the classical solution  is (cf. \red,\bou)
\eqn\ttt{\S _{R,L}(\t,\s) = e^{-{i\ov 2}  q y(\t,\s)  } \Sigma_{R,L} (\s_\mp)\
, \ \ \
 \Sigma _{R,L }(\t,\s + \pi ) = e^{ i \pi \g   } \Sigma_{R,L} (\t,\s)\ ,
  }
so that the change  $\g\to \g +1$  is equivalent to the change of sign in the
boundary conditions (in $\s$) for the free  fermion field
$\Sigma_{R,L}$.\foot{The Hamiltonian  contains  a term
 of the form (cf.  \kkk )
$\sum_{m} (m-\ha  \g) s_m^* s_m$, where $s_m, s^*_m$ are standard
GS fermion oscillators.
The vacuum state in Fock space  is defined in the standard way as being
annihilated by negative frequency oscillators, i.e. $s_m|0\rangle=0,
\ m>\ha \g$. In the interval $2k-1\leq  \gamma<2k$ we have
$\g -2k+1< 1$, so  that it is convenient to represent  this term as
 $
\sum_{m} [m-(\ha \g-k+\ha ) -k +\ha ] s_m^* s_m$  $=$ $
\sum_{r} [r -(\ha \g-k+\ha ) ] {s_r'}^* s_r' ,\ \ s'_r\equiv s_{m-k+\ha} .
$
This will   lead to  the
expressions for $\hat N_{L,R}$ and $\hat J_{L,R}$ in $\hat H$
of the type  which appear in the NS-sector in the RNS formalism,  i.e.  with
$-\ha$ normal-ordering constant.}
The fact that  for
$2k-1\leq  \gamma<2k$ the  GS operators $\hat N_{R,L}$
 take half-integer eigenvalues  indicates, in particular, that in these
intervals the GSO projection that must be done
 in the RNS approach must be the `reversed' one.

In general, the  model  with $qR=2n +1$ ($\g= 2k +1$ for odd $w$)
is equivalent to the free superstring compactified on
  a twisted 3-torus (in the limit when the 2-torus part is replaced by
2-plane), or on a circle with antiperiodic boundary conditions for the
fermions \rohm\ (in particular, the theory with $qR=1$ and
$R<\sqrt{2\a'}$ will
have  tachyons). 

The fundamental world-sheet fermions  $\S$ that
appear in GS action \gss\ are always {\it periodic}  in $\s$.
This is necessary for supersymmetry of the model in the $q=0$ 
limit. If one  considers  spinors (space-time fermions)
 in the space with the metric (2.6)
 one should assume 
that they  satisfy  the periodic boundary conditions in $y$
for  arbitrary $q$
since this is the condition of unbroken supersymmetry 
in the limit $q\to 0$.
This condition of correspondence with the standard superstring 
  `free-theory' limit 
fixes the ambiguity in the choice of a spin structure which a priori exists
for all $qR=m$. Then  the adequate  point of view is that the breaking of supersymmetry  for 
$qR\not=2n$ (in particular, for $qR=2n+1$) 
is due to the non-trivial background ($q$-dependent spin connection)
and not due to  `special' choice of boundary conditions.

The `redefined' fermions $\Sigma$ in \ttt\
are effectively dependent on $y$ and thus  change phase under
a shift in $y$-direction. For  $qR=2n +1$ this results in antiperiodic
boundary conditions for {\it space-time} fermions as functions of $y$
(the space-time fields can be represented, e.g.,    as coefficients in
expansion
of a super string field $\Phi (y,\S, ...)$
in powers of world-sheet fermions).
As a consequence, there exists a continuous 1-parameter family of models 
interpolating between  the  standard supersymmetric $qR=0$  model with  fermions which are periodic in $y$ and a non-supersymmetric  $qR=1$ model with fermions which are antiperiodic in $y$.

The GS formulation  makes also transparent
the computation of the 1-loop (torus)  partition function
(which  will be  non-vanishing  for $q\not=2nR\inv$
due to the absence of fermionic zero modes, i.e. the absence  of
supersymmetry).
Indeed,  the path integral computation of the  GS superstring
partition function is  a straightforward generalization of the
 computation  in the bosonic string model described in \rut.
The first step is to  expand $y$ in  eigen-values of the Laplacian on the
2-torus and redefine
the fields $x,x^*$ and $\S_{L,R}, \bar \S_{L,R}$ in \ongs,\jjj\ to eliminate
the non-zero-mode part of $y$ from the   $U(1)$ connection.
The zero-mode part  of $y$ on the torus
($ds^2 = |d \s_1 + \t d \s_2 |^2  , \ \ \t=\t_1 + i\t_2 , \ \
 0<\s_a\leq 1$) is  $y_* =y_0 +
 2\pi R(w \s_1 + w' \s_2)$,  where $w,w'$ are
integer  winding numbers. Integrating over the fields $x,x^*$ and $\S^r_{L,R},
\bar \S^r_{L,R}$,
we get a ratio of determinants of  scalar  operators  of the
type $\del + iA, \ \bd - i\bar A $ \ ($\del =\ha (\del_2 -\t
\del_1)$)    with constant connection
\eqn\yyt{  A=q\del y_* = \pi \chi , \   \
\bar A= q \bd y_* = \pi \bar \chi    , \ \ \
\    \chi\equiv    qR (w' -\t w) ,  \   \
\bar \chi\equiv  qR ( w' -\bar \t w)  .
}
The final expression for the partition function takes the simple form
 (cf. \rut)\foot{Note that in the light-cone gauge the free part
of the GS  superstring  measure is trivial
(up to  the $\t_2^{-6}$ factor related  to the  zero modes)
 since the bosonic and fermionic
determinants  of $8$ bosonic and $8_L + 8_R$ fermionic degrees
of freedom
 cancel out (see  \refs{\bri, \carl}).}
\eqn\zzz{  Z(R,  q) = c  V_7 R \int
 {d^2\t \ov  \tau_2^2 }  \sum_{w,w'=-\infty}^{\infty}
 \exp \big( - {\pi (\a' \t_2)\inv R^2 |w' -\t w|^2 } \big) \ } $$ \times \
{\cal
Z}_0 (\t, \bar \t;\chi,\bar \chi )    \  {Y^4 (\t, \bar \t;
\ha \chi ,\ha \bar \chi  ) \ov Y(\t, \bar \t;
 \chi , \bar \chi )}\  . $$
Here
\eqn\yy{ Y (\t,\bar \t; \h, \bar \h)\equiv
{{\det}' (\del +  i\pi  \chi) \ {\det}' (\bd - i\pi  \bar \chi)\ov
{\det}' \del \  {\det}' \bd } = {U(\t,\bar \t; \chi, \bar \h )\ov
U(\t,\bar \t; 0 , 0 ) } \ ,  }
 \eqn\yyy{
U(\t,\bar \t; \chi, \bar \h )
\equiv    \prod_{(n,n')\not=(0,0)}
 (n'- \t n + \chi )(n'-\bar \t n + \bar \h )
 \ ,  }
where, in  the determinants,  we have  projected out the zero modes appearing
at $\chi=\bar \chi=0$ ( i.e.   $Y (\t,\bar \t; 0, 0)=1$).
 The equivalent form  of $Y$ is  (see,  e.g.  \alvi\ and  \rut)
\eqn\ttyy{
Y (\t,\bar \t; \h, \bar \h) =
 \exp[{{\pi  (\chi-\bar \chi)^2
\ov 2 \t_2}}]  \     {\theta_1(\h| \t)
\ov \h\theta'_1 (0| \t) }  \  {\theta_1(\bar \h|\bar  \t)
\ov \bar \h\theta'_1 (0|\bar  \t) }  }
$$
= \
 \bigg|{  \theta \big[\matrix{\ha  + qRw \cr \ha  +  qRw' \cr} \big]  (0| \t)
\ov qR (w' - \t w) \theta'_1 (0| \t) } \bigg|^2  \ , $$
where $\theta_1(\h| \t) =
\theta \bigg[\matrix{\ha \cr  \ha \cr }\bigg]  (\h| \t)$.

The factor  ${\cal Z}_0$  in \zzz\  stands for
  the contributions of the integrals over  the constant  fields $x,x^*,
\S_{L,R}, \bar \S_{L,R}$  ( i.e.  the contributions of $(n,n')=(0,0)$ terms in
the determinants)
which become zero modes in the free-theory ($q=0$)
limit\foot{Note that the  full integrand of $Z$ is modular invariant  since the
transformation
of $\t$ can be combined with a redefinition of $w,w'$ (so that,  e.g.
 ${\cal Z}_0$  and $Y$ remain invariant).}
 \eqn\zerr{ {\cal Z}_0 = { (\ha \chi \t_2^{- 1/2})^4 \
(\ha \bar \chi \t_2^{- 1/2})^4 \ov \chi \bar \chi   \t_2^{-1} }\ ,
}
i.e. $ {\cal Z}_0 =2^{-8} q^6 R^6   |w' -\t w|^6  \t_2^{-3 } . $
$ {\cal Z}_0$  (and  thus $Z$)  vanishes for  $q\to 0$
in agreement with the restoration of supersymmetry (existence of fermionic zero
modes) in this  limit.\foot{The  $q\to 0$ divergence of the bosonic `constant
mode' factor $ \sim  q^{-2}$  corresponds to  the  restoration of the
translational invariance in the $x_1,x_2$-plane  in the zero magnetic field
limit
(this infrared divergence reproduces  the  factor of area of the 2-plane).
This factor was projected out in \rut\ to get a smooth $q\to 0$ limit
of $Z$.  As   is clear from the above, in the superstring theory
this divergence is
cancelled  against the   analogous  fermionic `zero-mode' factors
  which ensure the regular (zero) $q\to 0$ limit of $Z$.}

The partition function  vanishes at all points $qR=2n$  where the fermionic
determinants  have   zero modes  (or $\theta_1$-functions in $Y$-factors in
\zzz\ have zeros for any $w,w'$,\  $\theta_1(0| \t) =0$),
in agreement with the fact that the theory is trivial at these points. More
generally,  the theory, and,  in particular, $Z$  is periodic in $q$ (see
\inve)
 \eqn\peri{Z (R,q)= Z(R, q + 2nR^{-1})\ , \ \ \ \ \ n=0, \pm 1, ... \  .}
For $qR= 2n +1$ the partition function (with bosonic zero-mode singularity
properly regularized)
  is the same as that of
 free  superstring compactified
on a circle with antiperiodic boundary conditions
for space-time fermions \rohm\ (as was  already mentioned,
the dependence on odd $qR$  can be eliminated  from \jjj\
provided $\S_{R,L}, \bar \S_{R,L}$
satisfy antiperiodic boundary conditions in $y$).

Separating  contributions of different intervals of  values of
$w,w'$ in the sum in \zzz\  (which correspond to different values of $\g$  in
the Hamiltonian picture after Poisson
resummation)  and comparing with the RNS expression it can be confirmed
that the different  prescriptions for the GSO projection in the different
sectors discussed
above are consistent with modular invariance.

$Z$ is infrared-divergent for  those  values of the moduli
$q$ and $R$
for which there are tachyonic states in the spectrum (see Section 4)
and is finite for  all other values of
$q,R$ (a special symmetry of a general class of tachyon-free string models
 with finite 1-loop cosmological constant
was discussed in \diens).

\newsec{$a=1$ Melvin model and other more general static magnetic flux tube
models }
 In the previous Sections we  have discussed the simplest possible
static magnetic flux tube model.  More general  bosonic string
models  were constructed in \rut.
They  depend on 4 real parameters  $(R, q,\a,\beta)$, with
two  (`left' and `right') magnetic fields proportional to
$ q+\beta$  and $ q-\a $ and an antisymmetric tensor proportional to $\a-\beta
$.   The most interesting   subclass of  these  models,  corresponding  to the
$\a=\beta$ case,
  describes
 {\it static}  magnetic flux tube  backgrounds. It  contains
the $a=\sqrt 3$  Melvin model studied  above  as the special case of
$\a=\beta=0$   and the dilatonic $a=1$ Melvin model  as the case of
$\a=\beta=q$. The four-dimensional geometry is given by
\eqn\backg
{ ds^2_4=  - dt^2  + d\r^2 +   F(\r)\td F(\r)  \r^2
d\vp^2 + dx_3^2 \ ,\  
}
\eqn\baag{ {\cal A}_\vp=  q  F(\r)   \r^2 \ , \ \ \ \
{\cal B}_\vp=  -\beta \td F(\r)   \r^2 \ , \ }
$$ 
e^{2(\phi-\phi_0)}=\td F(\r )\ ,\ \   e^{2\s} = \td F(\r) { F}\inv
(\r) \ , \ \ \  F  \equiv {1\ov 1 +  q^2 \r^2}  \  ,\ \ \   
\td F  \equiv {1\ov 1 + \beta ^2 \r^2} \ .   
$$
The models with $\beta >q $ are
related to the models with $\beta<q$ by the  duality transformation in the
Kaluza-Klein  coordinate $y$; more precisely, the $(R, \beta,q)$ model
is  $y$-dual to $(\a'/R, q, \beta)$ model (so that  the $a=1$
  Melvin  model is  the `self-dual' point).
 For fixed $q$ these  models thus fill an interval $0 \leq \beta \leq q$
 parametrized by  $\beta$   with
$a=\sqrt 3$ and $a=1$ Melvin models  being the boundary points.
The non-trivial part of the corresponding Lagrangian is \rut\  (cf. \mee)
\eqn\lagg{
 L=
    \del_+ \r  \del_- \r
 +   F (\r)  \r^2
[   \del_+ \vp  +  (q +  \beta)    \del_+ y ]
 \  [  \del_- \vp + (q -\beta)  \del_- y ]   }
$$
  +  \  \del_+ y  \del_- y  +   {\cal R} (\p_0 +  \ha \ln F ) \ ,
\ \ \ \ \
  \  F\inv =   1 + \beta ^2  \r^2  \  . $$
This model is related to  the model \mee\ by the formal $O(2,2)$
duality rotation (combination of a shift of $\vp$ by $y$ and duality in $y$).
Indeed, it can be formally obtained from  the $y$-dual \duaa\  to \mee\
 by first changing $q\to \beta, \ \td y \to y$ in \duaa\ and then shifting $\vp
\to \vp + q y$. This explains  why this
bosonic model is solvable  even though the ten-dimensional target space
geometry is, in general,  no longer
flat.\foot{In \rut\ we used  the `rotating' coordinate system  by redefining
$ \vp \to  \vp  - \beta t \  $ (the corresponding  background remained
static).
This redefinition  is not actually necessary for the solution of the model as
we  shall explain below.}
The equivalent form of \lagg\ is
\eqn\lagge{
 L=
    \del_+ \r  \del_- \r
 +   F (\r)
[  \del_+ y -  \beta  \r^2  \del_+ \vp'  ]
  [  \del_- y +  \beta  \r^2 \del_- \vp'  ]   }
$$
  +\    \r^2  \del_+ \vp'  \del_- \vp'  +   {\cal R} (\p_0 +  \ha \ln F ) \ ,
$$
where  we have used the formal notation $\vp'= \vp + q y$. Introducing an
auxiliary 2d vector field with components $V_+, V_-$  we can represent  \lagge\
 as follows, cf. \cov\ (this corresponds to `undoing' the duality
transformation mentioned above)\eqn\lage{
 L=
 \ha ( \del_+ + i\b V_+ + iq \del_+ y )x \  ( \del_- - i\b  V_- -i q \del_- y
)x^*  + c.c. }
$$
+ \  V_+ V_-  -  V_-   \del_+ y  + V_+   \del_- y   \ .  $$
Now it is easy to understand why  the classical equations of this model are
explicitly solvable in terms of free fields
and  the partition function is computable.
In spite of the $y$-dependence in the first term, the equation of motion for
$y$  still imposes the constraint that $V$ has zero  field strength,
 ${\cal F}(V)=  \del_- V_+ -  \del_+  V_-=0$:
 the variation over $y$ of the first term vanishes once one uses the equation
for $x$ (this follows from the fact that $qy$-terms can be formally absorbed
into a phase of $x$). Then  $V_+ = C_+  +\del_+ \y, \  V_-= { C_-} +  \del_-
\y, \  C_\pm =\const$. In the equations  for $V_+, V_-$ one can again ignore
the variation of the first term in \lage\
since it  vanishes  under ${\cal F}(V)= 0$. We find that
$V_+= C_+ +  \del_+ \y=  \del_+ y, \  V_-= { C_-} +  \del_- \y= - \del_- y$.
The solution of the model then effectively reduces to that  of the model
\mee ,   the only extra non-trivial contribution being the zero mode parts  of
the two dual fields $y$ and $\y$.
 Interchanging  of $q$ and $\beta$ is  essentially equivalent (after solving
for $C_+,C_-$) to interchanging $y$ and $\y$
and thus momentum and winding modes.

Eliminating $C_+,C_-$
one gets terms quartic in the angular momentum operators in the final
Hamiltonian.
Similar approach applies to the computation of the partition function  $Z$.
Once $x,x^*$ have been integrated out, the integrals over the constant parts of
$V_+,  V_-$ cannot be easily computed  for $q\beta \not=0$
  and thus   remain in the final expression \rut\ (see also below).

This  discussion has  a
straightforward generalization to superstring  case.
 A simple way to obtain the  supersymmetric version of \lagg\  is to start with
\onn\ (with $\beta$ instead of $q$), make  the $y$-duality transformation,
\eqn\laggs{
 L_{\rm RNS}=  \del_+ x  \del_- x^*  +    \l^*_R  \del_+  \l_R  + \l^*_L
\del_- \l_L
} $$  +  F (x)  \big[\del_+ y  +{i\ov 2}
\b ( x \del_+ x^* - x^*  \del_+ x+
 2 \l^*_L  \l_L) \big]\big[\del_- y   - {i\ov 2} \b( x \del_- x^* - x^*  \del_-
x   + 2 \l^*_R  \l_R)   \big] $$ $$
   + \    {\cal R} (\p_0 +  \ha \ln F ) \ ,  \  \ \ \ \ \ \ \ F\inv = 1 + \b^2
xx^*\ ,
$$
 and then include the  $q$-dependence  by rotating $x$ and $\l$, i.e.   by
replacing their  derivatives  by covariant derivatives with $iq \del_\pm y$
  as a connection.
The action
now contains the  quartic fermionic terms  reflecting the non-trivial
(generalized) curvature of the space.
The model  still  remains solvable.
The direct analogue of
 \lage\  is  (cf. \onn)
\eqn\lages{
 L_{\rm RNS} =  \ha ( \del_+ + i\b V_+ + iq \del_+ y )x \  ( \del_- - i\b  V_-
-i q \del_- y )x^*
+ c.c. }
$$
+\   \l^*_R  ( \del_+ + i\b V_+ + iq \del_+ y )  \l_R  + \l^*_L  ( \del_- + i\b
 V_- + i q \del_- y)
  \l_L
$$
$$
+\   V_+ V_-  -  V_-   \del_+ y  + V_+   \del_- y   \ .
$$
 The final expressions for the Hamiltonian
and partition function then  look very similar to the bosonic ones
(the role of fermions is just to supersymmetrize the corresponding  free
superstring  number of states and  angular momentum operators  and to cancel
certain normal ordering terms).

 The operator quantization of the model can be performed in a similar
way as in the simplest  case of
 $a=\sqrt{3}$ Melvin model. The exact Hamiltonian corresponding to the
superstring theory on the 
curved space-time geometry \backg , \baag \  takes a very simple and
$\beta$-$q$ symmetric form, cf. \hamil\
\eqn\hail{
\hat H =
 \ha  \a' ( -E^2 +  p_\a^2  ) + \hat N_R+  \hat N_L }
$$ +\   \ha \a' R^{-2} (m- qR\hat J)^2
+ \ha {\a'}\inv R^2 (w  -  \a' \b R\inv  \hat J)^2
-    \hat \g  (\hat J_R-\hat J_L)\ , $$
\eqn\cos{\N_R-  \N_L = mw  \ ,   }
\eqn\gam{ \hat \g\equiv \g - [\g]\ ,\ \ \ \ \  \ \g\equiv  qRw + \a' \b R\inv m
- \a' q\b \hat J \ , }
where $[ \g]$ denotes the integer part of $ \g$ (so that $0\leq \hat\g<1$)
and the operators  $\hat N_{R,L}, \ \hat J_{R,L}$ are the same as in \hamil.

 The duality symmetry in the compact Kaluza-Klein direction $y$ (which
exchanges  the axial and vector magnetic field parameters  $\beta $ and $q $)
is now manifest. The Hamiltonian is indeed
invariant under  $R\leftrightarrow \a'R\inv ,\  \beta \leftrightarrow q
\ \ m\leftrightarrow w$.
The resulting expression for (mass)$^2$ is obvious from \hail\ (cf.
\kme,\masgr). The mass formula  can also be  written in terms of  the `left'
and `right' magnetic field parameters
and charges, $\  B_{L,R}\equiv  q\pm \b$, \ \  $Q_{L,R}= {mR\inv } \pm {
\a'}\inv Rw$,
\eqn\hamelv{
    \a' M^2   =  2\hat N_R +  2\hat N_L + \ha \a' (Q_L^2 + Q_R^2 )
 } $$ - \  2 \a'   \big(B_L  Q_L \hat J_R + B_R Q_R\hat J_L \big)
+ \a' \big(B_L^2 \hat J_R+B_R^2 \hat J_L\big) \hat J  \  .
$$
It  is clear from eq. \hail\ that  all states with
$\hat J_R-\hat J_L\leq \hat N_R +\hat N_L $ have positive mass squared.
The only bosonic states which can be tachyonic   thus lie on
the first Regge trajectory with maximal value for $S_R$, minimal value for
$S_L$,  and   zero orbital momentum,
i.e.  $\hat J_R=S_R-\ha=\hat N_R+ \ha ,\   $  $\hat J_L=S_L+\ha=-\hat N_L -\ha
$, so that $\hat J_R-\hat J_L= \hat N_R +\hat N_L +1$.
Then
\eqn\mmm{
\a' M^2   = 2(\hat N_R + \hat N_L)(1 -\hat \g)}   $$  + \
  \a' R^{-2} (m- qR\hat J)^2
+  {\a'}\inv R^2 (w  -  \a' \b R\inv  \hat J)^2
   - 2\hat \g  \ ,  $$
which is not positive definite due to the last term $-2\hat \g $.
For all other possible values of $\hat J_R , \ \hat J_L$  the resulting  $M^2$
is non-negative.   In particular, all fermionic  states
 will have  (mass)$^2\geq 0$, as expected in a unitary theory.
This is manifest from eq. \hail , except for the  fermions with
  $\hat J_R-\hat J_L= \hat N_R +\hat N_L +\ha$, for which
there is  a negative contribution
 $-\hat \g$  in the expression for $M^2$. A
close inspection of eq. \hail\ shows that  $M^2\geq 0$ is true also
 in this case.

{}From eq. \mmm\ one learns  that in general there are
instabilities (associated with states with  high spin and charge) for
arbitrarily small values
of the magnetic field parameters.
The special case of $\beta=0$ (or $q=0$),  corresponding to  the $a=\sqrt{3}$
Melvin model discussed in Section 4, is the only exception:  we have seen that
in this  (type II) model  there are no tachyons
 below some {\it finite} value of $q$.
Let us now consider an example which illustrates the generic pattern:  the
$a=1$ Melvin model
where   $q=\beta $ ($B_R=0, B_L= 2\b$) and
\eqn\hmelv{
    \a' M^2   =  4 \hat N_R + \a' Q_R^2 -4\hat \g  \hat J_R
\ , \ \ \  \ \  \g =\a' \b  Q_L - \a' \b ^2 \hat J\ .
}
Let us take  for simplicity $R=\sqrt{\a'}$,
and choose the  states with $w=m$, \    $\hat N_L=0$, $\hat J_ R=\hat N_R+ \ha
$ and
$\hat J_ L= - \ha $. These states
 become tachyonic for $\b$ in the interval
$\beta_{1}< \beta < \beta_{2}$, with
\eqn\ancr{
\b_ {1,2}={1\ov m} \big( 1\mp \sqrt {1- \g_{\rm cr} }\big) \ ,\ \
\ \ \ \ \  \g_{\rm cr}={m^2\ov m^2+\ha} \ .
}
For large $m$ these magnetic field parameters  will be very small. Conversely,
given any arbitrarily small magnetic field, there will be tachyons
corresponding
to  states with  $m$ obeying
$
 {\beta\inv }  - 2^{-1/2} <m<  { \beta\inv }+ 2^{-1/2},
$
where  we have neglected  $O(\beta)$ terms.
Unlike  the usual Yang-Mills type  magnetic instabilities,   these
(being associated with higher level states)
remain even after the massless level states get  small masses
(they can   be eliminated only if the corresponding  higher-spin states receive
Planck-order corrections to their free-theory masses).

For generic values of the magnetic field parameters $\beta, q$  the
supersymmetry is  broken in all these models. This can  be seen directly from
the spectrum.
Indeed,  the two  magnetic fields couple to  both components of the
spin ($S_L$ and $S_R$), which
cannot be simultaneously the same for bosons and fermions. This means  that
bosons and fermions  should  get  different  mass shifts.
When $qR=2n_1$  and $\a'\beta R\inv=  2n_2$ , $   \  n_{1,2}=0, \pm
1, ...,  $  the theory is equivalent to the free
superstring compactified on a circle (in this case  $\hat \g =0$ and,
after appropriate shifts of $m,w$ by  integers,  eq. \hail\
reduces to the  free superstring Hamiltonian). If $qR=2n_1+1$
or $\a'\beta R\inv =  2n_2+1 $, then the necessary shift in $m$ or $w$
in the fermionic sector
involves half-integer numbers. As discussed in the previous sections, in these
cases the theory can be interpreted as a free superstring  on a circle
 with antiperiodic
boundary conditions for space-time fermions.

Finally, the partition function can   be computed by a  similar procedure
as in  the bosonic case  \rut.
 Starting with the analogue of \lages\ in the Green-Schwarz approach
we  find (cf. \zzz)
\eqn\zzzh{  Z(R,  q, \b ) = c  V_7 R \int
 {d^2\t \ov  \tau_2^2 }  \int dC d\bar C  \ ( \a'  \t_2)\inv
 \sum_{w,w'=-\infty}^{\infty}
} $$
\times \exp\big( -  \pi (\a' \b ^2\t_2 )\inv
 [ \h \bar \h   - R (q + \b )  (w'-  \t w) \bar \h
 - R (  q -\b )      (w'- \bar \t w)\h $$
$$
+ R^2 q^2   (w'-  \t w)(w'- \bar \t w)]\ \big) \ $$
$$ \times  \ {\cal Z}_0 (\t, \bar \t;\chi,\bar \chi ) \ {Y^4 (\t, \bar \t;
\ha \chi ,\ha \bar \chi  ) \ov Y(\t, \bar \t;
 \chi , \bar \chi )}\
 , $$
$$
  \h   \equiv 2\b C +  qR(w'-\t w) \ , \ \  \
\ \ \ \bar \chi = 2\b \bar C  +  q R(w'-\bar \t w)    \ ,
$$
where $Y(\t, \bar \t; \chi , \bar \chi )$ and $ {\cal
Z}_0 (\t, \bar \t;\chi,\bar \chi )$ were defined in \ttyy\
and \zerr . The  auxiliary
parameters
  $C,\bar C$ are proportional to the constant parts of  $V_\pm$ in \lages.
 In the limit $\beta \to 0$  we recover the partition function  \zzz\ of  the
model discussed in the previous section.

The   partition function  \zzzh\   has the following  symmetries (cf. \peri),
\eqn\simm{Z(R,q,\beta)=Z(\a'R\inv, \beta, q) \ ,  } 
\eqn\siml{
Z (R,q,\beta)= Z(R, q + 2n_1 R^{-1},\beta +2n_2 {\a'}\inv R)\ , \ \ \ \ n_{1,2}
=0, \pm 1, ... \ .   }
These are  symmetries of the full conformal field theory (as can be seen
directly  from the string action in the Green-Schwarz formulation).
For $  qR  \neq n_1 $ and $\a'\beta R^{-1} \neq n_2$, $n_{1,2}=$ integers, there are tachyons at any value of the radius $R$, and the partition function contains infrared divergences. As follows from eq. \siml ,  when  $\a' \beta/R$  (or $qR$) is an even number,
the partition function reduces to that of the $a=\sqrt{3}$ model, eq. \zzz .
In particular,  in the special case   that both $qR$ and  $\a'\beta R^{-1} $
are even, the partition function is identically zero, since  for these 
values of the magnetic field parameters the theory is equivalent  to the free superstring theory. In the case when either $\a' \beta/R$  or $qR$ is an odd number, the partition function is finite in a certain range of values of the radius. 

\newsec{Conclusions }
The  simple model considered in the main part of  this paper  describes
  type II superstring  moving in  a flat  but topologically non-trivial
10-dimensional space.
The non-trivial  3-dimensional part of this space \met\  is a `twisted' product
 of a 2-plane and a circle $S^1$  (the periodic shifts in the   coordinate  of
$S^1$
being  accompanied  by rotations in the plane). The free continuous
moduli parameters are the radius  $R$ of $S^1$  and the `twist' $q$.
If    other 5 spatial
dimensions are toroidally compactified,  the model can be interpreted as
corresponding
 to the Kaluza-Klein Melvin magnetic flux tube background in 4 dimensions
($R$ being Kaluza-Klein radius and $q$ being proportional to the magnetic
field strength).

This   model  can  be  easily solved either
in the RNS or light-cone GS approach and
exhibits several  interesting features.
 The supersymmetry is broken if $qR\not=2n$.
For $qR=2n$ the theory is equivalent to the standard free superstring theory
compactified on a circle with periodic boundary conditions for space-time
fermions;   for $qR=2n+1$ it is equivalent to the free superstring with
antiperiodic boundary conditions for the  fermions (the model thus continuously
interpolates between these two free superstring models).
 The mass spectrum is invariant under $q\to q +2nR\inv$ and contains tachyonic
states for certain
intervals of values  of $R$ and  $q$.
 The  one-loop vacuum amplitude  $Z(R,q)$ is finite for $R>\sqrt{2\a'}$
 but diverges for those  $R$ and $q$,  for which
there   are  tachyonic states in the spectrum.

The presence of  tachyonic instabilities for certain finite values of $R$ and
$q$  is not surprising  in view of the magnetic interpretation of this model.
This  perturbative instability of  the  Kaluza-Klein Melvin  background  as a
solution of  superstring theory may be more serious than its
 potential non-perturbative instabilities  discussed  at field-theory  level
 in \gaun.\foot{For comparison,  the
 $S^1 \times$ (Minkowski space) Kaluza-Klein vacuum
is perturbatively stable but
 may be unstable at the
non-perturbative  level \wit. Let us note also that the perturbative
instability of the Melvin background
suggests that other, related,
 more general  solutions, such as the    Ernst geometry \dowo \
which asymptotically reduce to Melvin, are also  perturbatively
unstable at the superstring-theory level.}

We have seen  that  the  superstring versions of
more general static magnetic flux tube
 models of \rut\ (which depend on  compactification radius,
vector and axial magnetic field parameters $R, q$ and $\beta$)
have analogous properties.
 In particular, supersymmetry  is  broken for all of these
models (for generic values of $\beta, q$).
These more general
models  reduce to the  free  superstring theory  when
both $qR$ and $ \a' \beta  R\inv $ are  even
integers.
 The bosonic  string partition function  \rut\ has the following symmetries:
$Z(R,q,\beta)=Z(\a'R\inv, \beta, q)$ and
$Z (R,q,\beta)= Z(R, q + n_1 R^{-1},\beta +n_2 {\a'}\inv R) , \  n_{1,2}=0, \pm
1, ...\  $.  The same  symmetries  are  present also  in the superstring case,
with the replacement $n_{1,2}\to 2 n_{1,2}$ (the case of odd
integers $n_1,n_2$ is again equivalent to the theory with  antiperiodic
fermions).

A common feature of all these models is the appearance  of tachyonic
instabilities associated  with states on  the first Regge trajectory.
All other Regge trajectories are tachyon-free.
This should
 be  a universal feature  of all static   backgrounds in superstring  theory.
Indeed, this fact is related to unitarity (implying the  absence of `fermionic
tachyons').\foot{Since a  unitary tree-level $S$-matrix  should correspond to a
string field theory  with  a hermitian  action,
 the  `square' of hermitian  fermionic kinetic operator should be
positive  in any  background. This translates into positivity of $M^2$ for the
fermionic states in the case of static backgrounds.} The expression for
(mass)$^2$  depends on  the angular momentum operator.
If there were bosonic tachyons not only on the leading  Regge trajectory,
but  also on  the subleading one, then a
fermionic state  with the  `intermediate' value of the spin (but otherwise
the same quantum numbers)
would have $M^2 <0$. Since this is not allowed by unitarity,
in any unitary superstring
 model corresponding to a static background
 tachyonic states can only appear on the first (bosonic) Regge trajectory.

The breaking of  supersymmetry  in the  model \onn\    is  a consequence of an
incompatibility
between periodicity of space-time spinors in the compact Kaluza-Klein direction
$y$
and the presence of a mixing between $y$ and the  angular coordinate of
 2-plane (this mixing  produces a flat but globally non-trivial
connection in the
fermionic derivatives).
Replacing the 2-plane  by a {\it compact}  space  with a non-trivial  isometry
parametrized by a
coordinate $\theta$ and mixing $\theta$ with another compact internal
coordinate $y$, one may try to construct
a similar model in which supersymmetry is broken with  preservation of Lorentz
symmetry in the remaining  flat non-compact directions.
The simplest examples of such models are  string compactifications
on twisted tori (or, equivalently,  string analogues of the
`Scherk-Schwarz' \sche\  or `coordinate-dependent' compactifications)
\refs{\rohm,\sch,\koun,\kou}.
 Consider, e.g., the  3-torus
$(x_1,x_2,y)\equiv (x_1 + 2\pi  R' n_1, x_2+ 2\pi  R' n_2 ,y+ 2\pi R k)$
and twist it by imposing the condition that the shift by period in $y$
should be  accompanied by a rotation in the $(x_1,x_2)$-plane.
For a {\it finite } $ R'$ the only possible rotations are by angles
$\ha \pi n$,  i.e.  one may identify the points
$(\theta, y)=(\theta +  2\pi n + \ha \pi  k, y + 2\pi R k), \
\ \cot \theta = x_1/x_2$.
  The superstring theory  with  this  flat but non-trivial
3-space  as   (part of) the  internal space
was  considered   in \rohm\ (see also \refs{\sch,\kou})
 where it  was found that  such a twist of the torus
breaks  supersymmetry and  leads to  the
existence of tachyons for $R^2<2\a'$
and a
 finite  (for $R^2> 2\a'$)
non-vanishing partition function.
It is easy to see that the    $ R'\to \infty$ limit of the Rohm model
is  actually equivalent to the special  case
 $qR=  \four k $  of our model.
 The corresponding limits of the spectra and partition functions of the two
models indeed  agree (the case of  $k=4$  explicitly  considered in \rohm\ is
equivalent to the superstring compactified on a circle with antiperiodic
boundary
conditions
for the fermions).
Since in the present model  the 2-plane is non-compact and thus the twisting
angle $2\pi qR$ is arbitrary,
 this model   continuously  connects
   large $ R'$ limits of the  models of  \rohm\ with different values of
integer $k$.

Similar models with compact  {\it flat}
 internal spaces obtained
by  `twisting' tori  always have {\it discrete}
 allowed values of the twisting
parameter (a symmetry group of a  lattice   which generates a torus from $R^N$
is discrete).
It is of interest to study analogous `twistings'
 of models with compact {\it curved} internal spaces with isometries.
 For example, one may consider
  the $SU(2) \times U(1) $ WZW model
and  `twist' the product  by shifting
the two  isometric Euler angles  $\theta_L$ and $\theta_R$
of $SU(2)$  by  the coordinate  $y$   corresponding to $U(1)$,
 $\theta_L' = \theta_L + q_1 y$, $\theta_R' = \theta_R + q_2 y$
($q_1,q_2$ are  arbitrary
continuous twist parameters). The
   model with  $q_2=0$ was recently  discussed   in \kok.
The resulting action $I_{SU(2)} (\theta_L', \theta_R', \psi) + \int (\del y)^2
$
defines a  conformal theory  (locally  the 4-space  is still
$SU(2) \times U(1) $ group manifold).
The case of $q_1= q_2$ (or $q_1= -q_2$)  is  a compact analog
of the model \mee\  studied  in the present paper.
It is possible to show that supersymmetry is  broken
(in particular, there is no Killing spinors)
in this `compact' model for all values of the continuous 
parameters  $q_i\not= 2nR\inv$ \teee.
 This is not, however, in contradiction   with  the `no-go' theorem 
of ref. \bank. 
In the case of compactification on $SU(2) \times U(1) $ group space 
the supersymmetry is  broken (in a `discrete' way)
already 
in the absence of twisting ($q_i=0$)  
due to the  central charge  deficit (see, e.g., \refs{\barny,\deal}). 
Still, analogous   closed string 
models  containing extra  continuous 
supersymmetry-breaking  `magnetic' parameters 
 may be of interest in connection with  a possibility of 
 spontaneous tree-level supersymmetry breaking in
string theory.

\vskip 1cm
 \noindent {\bf Acknowledgements}

\noindent
We are grateful  to L. Alvarez-Gaum\' e, M. Green  and  C. Klim\v c\'\i k  for
helpful  discussions.
A.A.T.   acknowledges the  hospitality of CERN Theory  Division   and the
support
of PPARC, ECC grant SC1$^*$-CT92-0789
and NATO grant CRG 940870.

\vfill\eject
  \listrefs
\vfill\eject
\end